\documentclass[pdflatex,sn-mathphys]{sn-jnl}

\usepackage{threeparttable}
\usepackage{amsmath}
\usepackage{url}
\usepackage{multirow}
\usepackage{lineno}
\jyear{2021}%
\theoremstyle{thmstyleone}%
\theoremstyle{thmstyletwo}%

\theoremstyle{thmstylethree}%

\begin{document}


\title[L. Bian et al.]{Towards Large-scale Single-shot Millimeter-wave Imaging for Low-cost Security Inspection}

\author*[1,2]{Liheng Bian }\email{bian@bit.edu.cn}
\equalcont{These authors contributed equally to this work.}
\author[1]{Daoyu Li}
\equalcont{These authors contributed equally to this work.}
\author[3]{Shuoguang Wang}
\equalcont{These authors contributed equally to this work.}
\author[3]{Chunyang Teng}
\author[3]{Huteng Liu}
\author[1]{Hanwen Xu}
\author[1]{Xuyang Chang}
\author[3,4]{Guoqiang Zhao}
\author*[3,4]{Shiyong Li}\email{lisy\_98@bit.edu.cn}
\author*[1]{Jun Zhang}\email{zhjun@bit.edu.cn}


\affil[1]{\orgdiv{MIIT Key Laboratory of Complex-field Intelligent Sensing}, \orgname{Beijing Institute of Technology}, \orgaddress{\postcode{100081}, \state{Beijing}, \country{China}}} 

\affil[2]{\orgdiv{Yanotze Delta Region Academy of Beijing Institute of Technology (Jiaxing)}, \orgaddress{\postcode{314019}, \state{Jiaxing}, \country{China}}}

\affil[3]{\orgdiv{Beijing Key Laboratory of
Millimeter Wave and Terahertz Technology}, \orgname{Beijing Institute of Technology}, \orgaddress{\postcode{100081}, \state{Beijing}, \country{China}}}

\affil[4]{\orgdiv{Tangshan Research Institute of Beijing Institute of Technology}, \orgaddress{\postcode{063007}, \state{Tanshan}, \country{China}}}




\abstract{Millimeter-wave (MMW) imaging is emerging as a promising technique for safe security inspection. It achieves a delicate balance between imaging resolution, penetrability and human safety, resulting in higher resolution compared to low-frequency microwave, stronger penetrability compared to visible light, and stronger safety compared to X-ray. Despite recent advances in the last decades, the high cost of requisite large-scale antenna arrays hinders the widespread adoption of MMW imaging in practice. To tackle this challenge, we report a large-scale single-shot MMW imaging framework using a sparse antenna array, achieving low-cost but high-fidelity security inspection under an interpretable learning scheme. We first collected 1934 full-sampled MMW echoes that are enough to study the statistical ranking of each element in the large-scale array. These elements are then sampled based on the ranking, building the experimentally optimal sparse sampling strategy that reduces the cost of the antenna array by up to one order of magnitude. Additionally, we derived an untrained interpretable learning scheme, which realizes robust and accurate image reconstruction from sparsely sampled echoes. Last, we developed a neural network for automatic object detection, and experimentally demonstrated successful detection of concealed centimeter-sized targets using 10$\%$ sparse array, whereas all the other contemporary approaches failed at the same sample sampling ratio. The performance of the reported technique presents higher than 50$\%$ superiority over existing MMW imaging schemes on various metrics including precision, recall, and mAP50. With such strong detection ability and order-of-magnitude cost reduction, we anticipate that this technique provides a practical way for large-scale single-shot MMW imaging, and could advocate its further practical applications.
}

\maketitle

\section{Introduction}\label{sec:introduction}


Security check at public places such as airports and railway stations requires effective personnel surveillance techniques to prevent rising-concerned terrorist attacks\cite{triplett2001technology}. However, conventional surveillance techniques have limited utility in practice, as metal detectors can only detect metallic weapons and explosives, X-ray machines expose individuals to harmful ionizing radiation\cite{li2020cylindrical}, infrared imaging systems\cite{scribner1991infrared,de2019real} are sensitive to environmental disturbance, and visible-light cameras cannot penetrate clothing. A practical security check system typically relies on the combination of these techniques. For example, an airport security screening process involves utilizing X-ray scans to inspect luggage and using metal detectors or manual methods to check if individuals are carrying prohibited items on their person. Consequently, the security system suffers from complex surveillance system design, additional security personnel and inspection workload, and poor efficiency of passage caused by separate screening of individuals and luggage. Thus, there is an urgent need for a security inspection approach that can simultaneously detect concealed items carried by individuals as well as those carried in luggage. In this context,  the millimeter-wave (MMW) imaging has emerged as a promising alternative, due to its unique advantages of high-resolution and penetrable imaging ability while being safer for human exposure compared to X-ray machines\cite{li2019machine}. MMW imaging provides the possibility of an all-in-one high-throughput security screening system with high resolution, high penetrability, and strong safety.

MMW imaging systems are generally categorized into two types: passive and active. Passive systems\cite{lynch2008passive} rely on detecting the naturally occurring MMW radiation emitted by the target. However, the imaging quality of passive systems is commonly limited due to the target's weak MMW radiation, which renders the system vulnerable to environmental noise and interference. Furthermore, they are limited to 2D imaging and hard to obtain targets' 3D information.
Active systems\cite{li2019machine,sheen2001three,liu2022programmable,hunt2013metamaterial,cui2014coding,li2017electromagnetic}, on the other hand, use extra MMW illumination and obtain scattered EM waves to achieve higher resolution imaging than passive systems. There are two widely adopted active MMW imaging frameworks, including single-input-single-output (SISO) and multiple-input multiple-output (MIMO) systems\cite{zhuge2012three}.
While a full SISO array can achieve high-quality imaging, it requires a large number of antenna elements that are prohibitively expensive to manufacture. As an illustration, considering a typical Ka-band security imaging system with a resolution of approximately 5 mm, it requires an antenna array on the order of $10^4$. The cost of such an array exceeds more than one million dollars. Consequently, a mechanically scanning linear array with range migration algorithm (RMA)\cite{sheen2001three} has emerged as a more economical alternative for MMW imaging. However, the resulting images may suffer from blur even if the subjects jiggle during data acquisition\cite{liu2023millimeter}, which degrades the performance of subsequent concealed object detection\cite{kupyn2018deblurgan}. Electrical scanning MIMO arrays demonstrate the advantage of requiring fewer elements while enabling data collection through a single snapshot. Nonetheless, current high-accuracy MIMO reconstruction methods entail either time-consuming coherent accumulation across the entire aperture\cite{desai1992convolution,zhuge2012three} or multistatic-to-monostatic transformation \cite{moulder2016development}. This leads to an inherent trade-off between image quality and running efficiency, posing a significant challenge\cite{fromenteze2019transverse,alvarez2014fourier,abbasi2018fast,li2021efficient,li2020cylindrical}.

Here, we report a novel scheme of low-cost large-scale MMW imaging for single-shot human security inspection, enabling successful detection of concealed centimeter-sized targets using a sparse array.
The differences between the reported technique and the conventional approaches lie in the following aspects. First, instead of regularly arranged or scanning arrays, we designed a sparse-optimized antenna array for single-shot MMW imaging. Specifically, we first collected a set of real-captured echoes using a full-sampled antenna array, and then analyzed the statistical importance ranking map of array elements. We theoretically proved that a larger value in the map indicates a more significant element in the antenna array. Based on the ranking map, we experimentally derived a statistically optimized sampling strategy to sparsely select antenna elements at various sampling ratios. Under low sampling ratios of such as 25$\%$, the technique experimentally exhibits 22$\%$ improvement on RMSE, 2.25 dB on PSNR, and 0.315 on SSIM compared to random sampling strategy.
This statistical approach allows for a more efficient selection of the most significant elements, thereby reducing the number of inefficient elements in antenna array design.

Besides, we introduced an interpretable untrained learning approach that enables robust MMW  reconstruction from sparsely sampled echoes. We optimize a lightweight complex-valued network by the objective based on the physical model of MMW scattering. This physics-informed learning strategy does not require any training, breaking the black-box limitation of training-based techniques. Compared to the existing CS-based approaches, it yields high-fidelity reconstruction with up to 3 dB improvement on PSNR at low sampling ratios. Besides, the technique shows prominent priority over existing techniques for detecting concealed targets from sparsely sampled echoes. 
The combination of statistics-based sparse sampling and untrained learning techniques results in a significant reduction in the cost of the antenna array by up to an order of magnitude, without compromising the system's ability to detect centimeter-sized concealed targets.

Furthermore, we employed a neural network for automatic detection of concealed targets. The detection network was trained on full-sampled reconstructions of the collected MMW imaging dataset to classify five kinds of common hidden targets, as opposed to the recent trend of single-class detection works\cite{Liu2019Concealed,Wang2022Self}. We evaluated the detection results with the sparsely sampled reconstructed images as input. The reported framework combining statistical sparse array and untrained reconstruction shows closer detection accuracy to the full-sampled detection results than existing approaches, which indicates a superior reconstruction accuracy. Experiments demonstrated that the reported imaging framework achieved successful concealed target detection at 10$\%$ sampling ratio, while other contemporary approaches failed. This work leverages the statistical prior and emerging fusion frameworks of neural networks and optimization, providing new insights into the development of low-cost and efficient MMW imaging systems.


\section{Results}\label{sec:result}

\subsection{Statistically optimized sparse array}
\begin{figure}[htbp]
	\centering
	\includegraphics[width=1\linewidth]{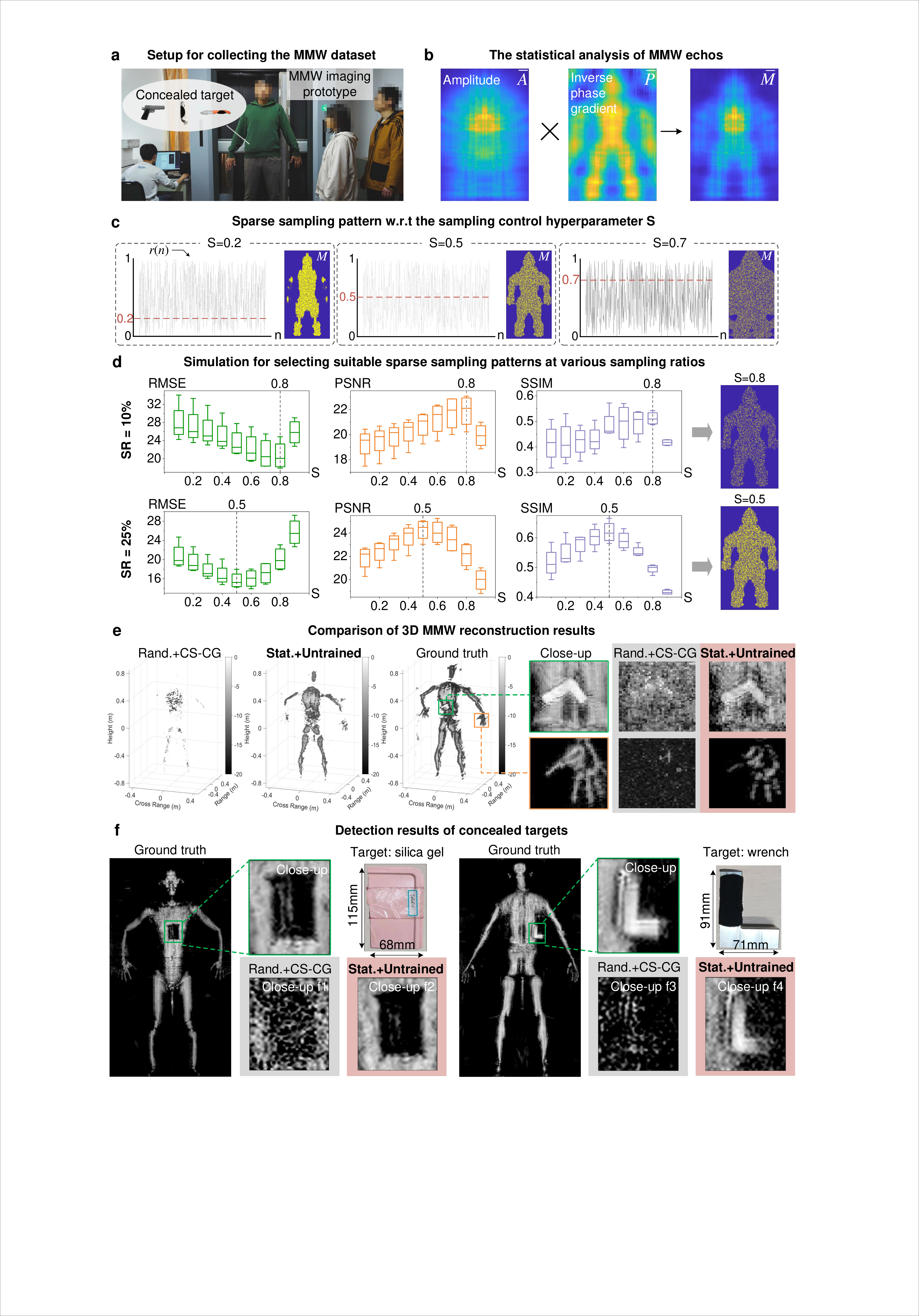}
	\caption{Principle of statistical sparse sampling. \textbf{a}, We collect a full-sampled MMW concealed target inspection dataset containing 1934 echoes. \textbf{b}, The statistical ranking $\bar{M}$ is obtained by multiplying the average amplitude $\bar{A}$ and inverse phase gradient $\bar{P}$. \textbf{c}, With the preset sampling control hyperparameter $S$ and a uniform random function $r(n)$ , we sample the $n$-th point if $r(n)>S$ following the statistical ranking in \textbf{b}. \textbf{d}, The evaluated RMSE($\downarrow$)/PSNR($\uparrow$)/SSIM($\uparrow$) curves of reconstructed 2D maximum value projections of randomly selected echoes w.r.t. different $S$. The optimal sampling masks of sampling ratio(SR)=$10\% \& 25\%$ correspond to $S=0.8\&0.5$. \textbf{e}, The reconstructed 3D scenes at 25$\%$ sampling ratio. \textbf{f}, The reconstructed results for concealed target detection at 25$\%$ sampling ratio. We compared the random sampling with conventional CS-CG\cite{li2018compressive} reconstruction and the reported statistically sparse sampling with the untrained reconstruction in \textbf{e} and \textbf{f}.}
	\label{fig:sample}
\end{figure}

In contrast to prevailing handicraft-designed arrays, our approach capitalizes on the statistical prior afforded by a large-scale dataset. To this end, we first gathered real echoes using a full-sampled antenna array. Specifically, we recruited 10 volunteers with various concealed objects, such as knives, pistols, scissors, metal strips, and explosive powder (using silica gel instead in experiments), and positioned them 0.3 meters in front of the array, as depicted in Fig. \ref{fig:sample} \textbf{a}. Throughout the data collection process, these volunteers were instructed to maintain a steady position. Ultimately, we obtained a total of 1934 echo samples, constituting the MMW imaging dataset. The dataset comprises samples with three dimensions: vertical, horizontal, and frequency, each containing $430 \times 186 \times 50$ voxels. To analyze the distribution of 3D echo, we extracted a 2D slice of echo at the center frequency along the frequency dimension. This particular slice was chosen as it provides the most comprehensive representation of the echo characteristics within this frequency band. We then calculated the average data of all the obtained 2D slices to generate a 2D average echo, which reflects the characteristics of the 3D averaged echo and facilitates investigation into the statistical importance ranking of elements in the 2D antenna array. Further, we extracted the average amplitude map $\bar{A}$ and inverse phase gradient map $\bar{P}$ (the reciprocal of phase's gradient) of the cross-section, as shown in Fig. \ref{fig:sample} \textbf{b}.
A larger value in $\bar{A}$ indicates that this element has more significance in the antenna array, and vice versa. As for the phase item, a continuous target distribution can result in a flattened phase gradient distribution of the array (see Sec. \ref{sec:method_gradient} and Supplementary Note 1). Consequently, we suggest that attention should be focused on the portion of the array with a smaller phase gradient distribution (larger values in $\bar{P}$). A larger value in the product of $\bar{A}$ and $\bar{P}$, denoted as $\bar{M}=\bar{A} \times \bar{P}$, signifies greater importance of the element within the antenna array. We set $n=1\to N$ as orders of each sorted element in $\bar{M}$. A smaller $n$ means higher statistical importance. Given the statistical importance ranking $\bar{M}$ and a uniform random function $r\left(n\right)$ ranging from 0 to 1 (Fig. \ref{fig:sample} \textbf{c}), we sample the elements $n$ of $r\left(n\right) > S$, where $S$ is a hyperparameter to control the sparsity of sampled pattern (Details in Sec. \ref{sec:method_sampling}).

We conducted a series of simulations to reveal the performance of the reported sparse sampling. In the following evaluations, the root mean square error (RMSE), peak signal-to-noise ratio (PSNR), and structural similarity index (SSIM)\cite{wang2004image} are employed to quantify reconstruction accuracy. We compared both the 2D maximum value projections along the range direction and 3D results reconstructed by various imaging algorithms. The dynamic range of the reconstruction results is set to 20 dB. Images retrieved by RMA from full sampling echoes are regarded as references. Figure \ref{fig:sample} \textbf{d} demonstrates there exists an optimal sparsity for a certain sampling ratio. For instance, the hyperparameter $S=0.8$ corresponds to the optimal sparsity of $10\%$ sampling ratio while $S=0.5$ corresponds to $25\%$ sampling ratio. The statistics in Tab. \ref{tab:diff_metrics_all} indicate that compared to the random sampling method, the statistically sparse sampling shows average improvements of 17$\%$ in RMSE, 1.7 dB in PSNR, and 0.24 in SSIM for CS-based reconstructions. Additionally, the advantages of the statistically sparse sampling become more pronounced at lower sampling ratio scenarios, with average improvements of 22$\%$ in RMSE, 2.25 dB in PSNR, and 0.315 in SSIM observed at a sampling ratio of 10$\%$.
Both visual and quantitative comparisons (Fig.\ref{fig:sample} \textbf{e} \& \textbf{f} and Tab. \ref{tab:diff_metrics_all}) validated the superiority of the reported statistically sparse sampling on reconstruction accuracy.

\subsection{The reconstruction results of untrained learning}

\begin{figure}[h!]
\centering
\includegraphics[width=1\linewidth]{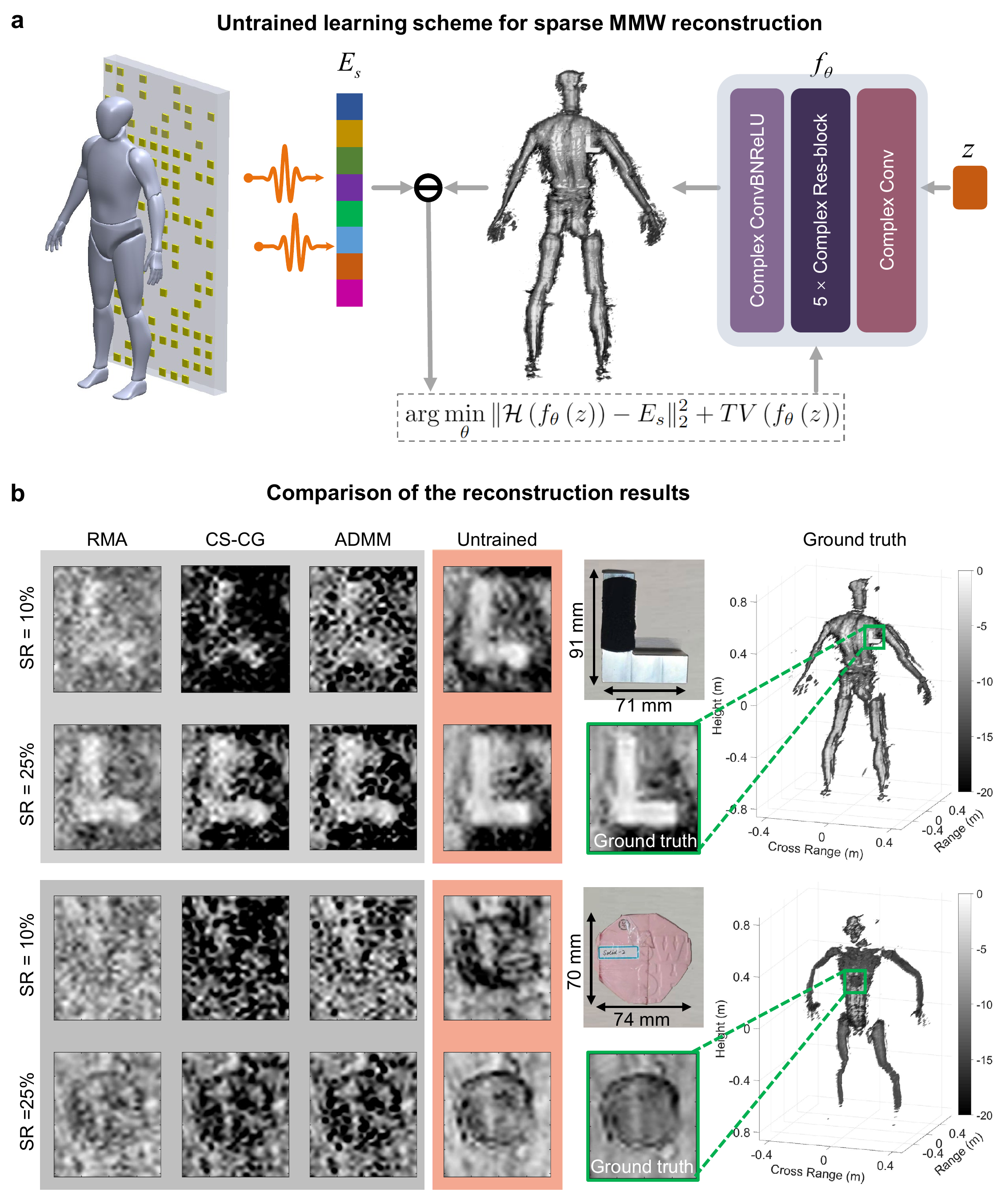}
\caption{The diagram of the reported lightweight complex-valued convolutional network (CCN). \textbf{a}, We optimize the parameter of CCN following the physical constraint (Eq. (\ref{eq:obj})) in an untrained manner to reconstruct the target scene. The CCN network is composed of a complex ConvBNReLU layer, five complex Res-blocks, and a complex Conv layer. \textbf{b}, The reconstruction results of various algorithms with the statistically sparse sampling.}
\label{fig:network}
\end{figure}

To ensure the fidelity of the reconstructed scene, we employ the physical prior to regulate the output of the neural network. We follow the interpretable untrained learning strategy\cite{ulyanov2018deep,wang2020phase}. By updating the parameters $\theta$ of the neural network $f_{\theta}$, the untrained reconstruction minimizes the loss between $\mathcal{H}\left(f_{\theta} \left( z \right)\right)$ and the measurement $E_{s}$, where $\mathcal{H}$ is the physical model of MMW scattering, and $z$ is the input of the network. Considering the complex-valued characteristics of the MMW echo, we construct a lightweight complex convolutional network (CCN)\cite{hirose2013complex,zhang2017complex,trabelsi2018deep} as shown in Fig. \ref{fig:network} \textbf{a}. Details can be found in Sec. \ref{sec:untrained}.
We apply the complex total variation (TV)\cite{gaoiterative} to enhance the output of network. The network does not require pre-training. Instead, it is the interplay between $\mathcal{H}$ and $f_{\theta}$ that causes the prior of $E_{s}$ to be captured by the neural network. When the optimization converges, the resulting $f_{\theta}$ is the inverse model of $\mathcal{H}$ and can be applied to reconstruct the target scene $I = f_{\theta}\left( z \right)$.

To evaluate the performance of the reported reconstruction, the common-used RMA and state-of-the-art CS-based methods including Compressed Sensing-Complex Gradient (CS-CG)\cite{li2018compressive},  Alternating Direction Method of Multipliers (ADMM)\cite{boyd2011distributed} are employed in the following comparisons.
As shown in Fig. \ref{fig:network}, the reported untrained reconstruction can retrieve more clear textures of concealed targets while RMA causes more artifacts, and other CS-based methods cannot mitigate the clutter components and maintain the details of the target. Moreover, the comparison in Tab. \ref{tab:diff_metrics_all} leads to the conclusion that the reported untrained learning outperforms existing algorithms in accuracy under both random and statistics-based sampling.  It shows about an average 3dB improvement in PSNR compared to existing CS-based approaches at low sampling ratios (10$\%$ and 25$\%$).

\begin{table*}[t]
\centering
\caption{The quantitative evaluations of reconstructed data with different sampling strategies and reconstruction methods.}	
\setlength{\tabcolsep}{3pt}
	\begin{threeparttable}\footnotesize
		\begin{tabular}{cc| cccc| cccc}
	\hline\hline
			
			\multicolumn{2}{c|}{Sampling strategy} & \multicolumn{4}{c|}{Rand.} &\multicolumn{4}{c}{Stat.} \\[0.3ex]
   \hline
		\multicolumn{1}{c|}{SR} & Metrics &RMA &CS-CG & ADMM & \textbf{Untrained} &RMA &CS-CG & ADMM & \textbf{Untrained}\\[0.3ex]
  \hline
  \multirow{3}{*}{10\%} & \multicolumn{1}{|c|}{RMSE($\downarrow$)} & 57.32 & 72.07 & 48.16 & \textbf{41.17} & 50.38 & 51.09 & 42.07 & \textbf{31.15} \\[0.3ex]
& \multicolumn{1}{|c|}{PSNR($\uparrow$)}& 13.03 & 10.99 & 14.55 & \textbf{15.94} &  14.15 & 14.11 & 15.75 & \textbf{18.38} \\[0.3ex]
 & \multicolumn{1}{|c|}{SSIM($\uparrow$)} & 0.161 & 0.031 & 0.238 & \textbf{0.474} & 0.250 & 0.537 & 0.529 & \textbf{0.623} \\ [0.3ex]
			\hline
  \multirow{3}{*}{25\%} & \multicolumn{1}{|c|}{RMSE($\downarrow$)} &  45.36 & 45.25 & 42.91 & \textbf{27.36} & 45.91 & 44.01 & 35.94 & \textbf{22.83} \\[0.3ex]
& \multicolumn{1}{|c|}{PSNR($\uparrow$)}& 15.07 & 15.13 & 15.65 & \textbf{19.48} & 15.16 & 15.39 & 17.24 & \textbf{21.04}\\[0.3ex]
 & \multicolumn{1}{|c|}{SSIM($\uparrow$)} & 0.261 & 0.391 & \textbf{0.607} & 0.588 & 0.301 & 0.640 & 0.685 & \textbf{0.764}\\ [0.3ex]
			\hline
		\end{tabular}
		\label{tab:diff_metrics_all}
	\end{threeparttable}
\end{table*}

The algorithms including RMA, CS-CG, ADMM, and the proposed network are executed on different platforms. Therefore, a direct comparison of their running time is not very meaningful. However, based on the current implementation, the average solution times of RMA, CS-CG, and ADMM methods running by Matlab on 2 $\times$ Intel E5-2687W CPUs are 2s, 610s, and  66s, respectively. The average running time of 100 iterations of the untrained method implemented by Pytorch on an NVIDIA RTX 3090 GPU is 34s. In most cases, it is enough to detect the targets in 80 iterations (less than 30s). The reported framework does not require a scanning process, so it can be adapted to most environments. This technique can be utilized in scenes with high-security requirements, such as airports and key train stations, where it can safely and conveniently inspect hidden items through clothing and enhance the efficiency of security checks.




\begin{figure}[h!]
    \centering
    \includegraphics[width=1\linewidth]{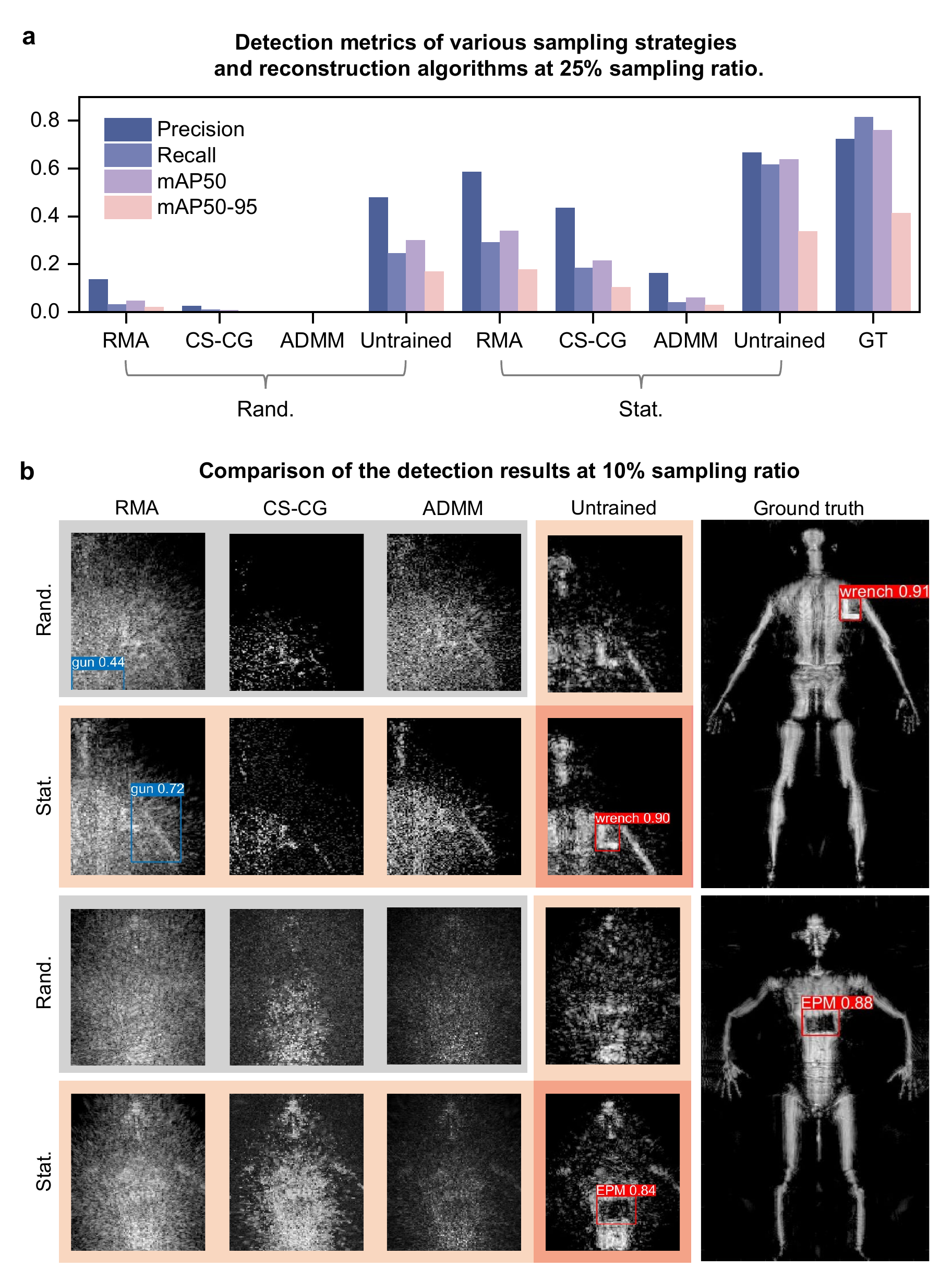}
    \caption{Detection results of various sampling strategies and reconstruction algorithms. We trained a YOLOv8 detector for multi-class concealed target detection.   \textbf{a}, The numerical comparison of detection performance under 25$\%$ sampling ratio. \textbf{b}, The comparison of the detetcion results at 10$\%$ sampling ratio. `EPM' is the abbreviation for explosive powdered material. Only the combination of statistical sampling and untrained reconstruction can stably recover image textures from 10$\%$ sparsely sampled echoes and detect the concealed targets with fairly high confidence coefficient.}
    \label{fig:min_sampling_ratio}
\end{figure}

\subsection{Concealed target detection results}
To evaluate the performance of the reported framework on post-processing concealed target detection, we adopted the commonly used object detection network YOLOv8\cite{YOLOv8} (see Supplementary Note 2). It is trained using full-sampled RMA images. Different from recent single-class detection works\cite{Liu2019Concealed,Wang2022Self}, the detection network is employed to distinguish five kinds of common kindden targets, such as knife, wrench, phone, gun, explosive powdered material (EPM, using silica gel instead in experiments), etc. We utilized commonly used indicators, including precision, recall, and mean average precision (mAP)\cite{zheng2015scalable}, to evaluate the detection results and assess the imaging methods' performance. The results are presented in Figure \ref{fig:min_sampling_ratio}. Although ADMM yields good performance in terms of image reconstruction, the reconstructed images are not suitable for concealed weapon detection, as shown in Figure \ref{fig:min_sampling_ratio} \textbf{a}. As depicted in Figure \ref{fig:network} \textbf{b}, the objects in the ADMM images exhibit fluctuating and perforated surfaces, which may impede the detection process. With a sampling ratio of 25$\%$, statistical sampling consistently outperforms random sampling in terms of detection accuracy. Furthermore, the untrained reconstruction method exhibits superior accuracy under both sampling schemes. The optimal performance among these techniques is attained by the combination of statistical sampling and untrained reconstruction at 25$\%$ sampling ratio with above 0.5 absolute numerical improvement on precision, recall, and mAP50 compared to the existing imaging schemes (Fig. \ref{fig:min_sampling_ratio} \textbf{a} and Tab. S1). 
Figure \ref{fig:min_sampling_ratio} \textbf{b} illustrates the detection results with bounding boxes and confidence coefficients at 10$\%$ sampling ratio. Under this extremely low sampling scenario, only the combination of statistical sampling and untrained reconstruction can yield accurate location and classification of the concealed objects with high confidence. Specifically, by employing statistical sparse sampling for antenna design and untrained learning for robust reconstruction, we can achieve successful single-shot detection of targets using only 10$\%$ of the original full antenna array elements. It indicates that this architecture is promising to reduce the cost of antenna arrays for single-shot MMW security inspection by an order of magnitude.

\section{Conclusion and discussion}
In this work, we developed a large-scale single-shot MMW imaging system for low-cost security inspection.  
We collected a large-scale MMW human security inspection dataset of real-captured MMW echoes. Based on the dataset, we proposed a statistical sampling approach for sparse MMW array design that maintains high-fidelity reconstruction quality while minimizing the number of elements. Besides, we report an untrained reconstruction technique based on a lightweight complex-valued neural network for sparsely sampled measurements. By integrating iterative optimization and neural network techniques, this approach overcomes the black-box limitation of end-to-end neural networks and offers superior reconstruction accuracy and high reliability for sensitive security inspection. Furthermore, we trained a concealed target detection network, achieving automatic high-precision detection using the sparsely sampled measurements. Experiments demonstrated that the reported framework achieved accurate detection of concealed centimeter-sized targets with a 10$\%$ sparsely sampled array, whereas other contemporary approaches failed at the same sampling ratio. Overall, the technique maintains the potential for decreasing more than one-order-of-magnitude cost for efficient MMW surveillance.

At present, there exists a gap between our prototype and real applications. First, all echo data in the collected dataset were obtained from detecting 10 adults (including males and females) whose heights ranged from 165 to 185 cm and weights ranged from 50 to 80 kg. However, it cannot meet the requirements for serving people with various ages and body shapes. Second, although the reported reconstruction technique is far more efficient than most of the existing CS-based algorithms, it still needs about 30 seconds to image the target scene. It meets the demand of normal security checks but is insufficient for high-throughput real-time detection. Last, we have outlined the superiority of the reported statistical sparse array and untrained reconstruction over existing approaches, but we have not performed targeted optimization on the subsequent detection network. Consequently, the current network's detection accuracy warrants further improvement, indicating the potential for enhancing the system's detection performance.

There are several directions for future work to break the above limitations. 1) It is crucial to improve the data diversity by collecting echo data corresponding to a variety of somatotypes. Besides, the sampling combining statistics-based and hand-craft design may contribute to handling corner cases. 2) The convolution-accelerate hardware such as GPU with higher FLOPs, FPGA\cite{rahman2016efficient,zhang2018caffeine} and specific SOC\cite{hegde2017caffepresso,meloni2018neuraghe} can be utilized in improving the speed of reconstruction in practice. Moreover, low-precision networks have been proven to increase efficiency with comparable performance\cite{hubara2017quantized,NEURIPS2020_13b91943}. 3) According to the subject’s physique priori, defining a region of interest (ROI) and ignoring the reconstruction outside ROI can effectively reduce the number of parameters in the network, thereby improving computational efficiency.
4) To further boost the efficiency and accuracy of the reported method, we can employ joint learning for both the sampling mask and recovery network on a large dataset. When the training converges, the network inputs sparsely sampled signals to output the reconstruction results without the need for additional iterations. 


More broadly speaking, the statistical sparse sampling can not only be utilized in SISO arrays but also in MIMO ones with minor changes. 
For topology design, we can apply the statistical sampling strategy to the equivalent SISO array of a full MIMO array. Then, the statistically sampled equivalent SISO array is mapped to the MIMO array with fewer antenna elements. 
We anticipate that the proposed technique enables new possibilities for achieving lower-cost and higher-throughput MMW security inspection.

Furthermore, the statistical sparse sampling approach is not limited to the aforementioned Synthetic Aperture Radar (SAR) imaging systems, as it can also be extended to electrical scanning imaging systems, including electronically scanned antennas (ESAs). In the context of dynamic beamforming techniques, conventional ESAs rely on phased array technology which enables high-fidelity beam control. However, the use of phased arrays is associated with significant system costs and complex hardware complexity, which limits their applicability in certain contexts. By employing the statistical sparse sampling strategy, it is possible to reduce the density of the array, resulting in significant cost savings. This approach is therefore a promising alternative to conventional dynamic beamforming techniques, particularly in scenarios where cost and hardware complexity are critical considerations.


An alternative approach to using expensive transceiver modules in ESAs is to employ artificial surfaces designed using metamaterials and holography\cite{li2019machine,li2022intelligent,hunt2013metamaterial,cui2014coding,li2017electromagnetic}. Such surfaces can be engineered to synthesize a desired radiation pattern, and can effectively convert a surface wave into free-space radiation\cite{minatti2016synthesis}. These reconfigurable arrays, which include reflection and transmission arrays, offer a promising alternative to traditional dynamic beamforming techniques. Commercial products such as the Eqo MMW security imaging system\cite{corredoura2006millimeter}, developed by Smiths Detection in collaboration with Agilent, employ a reconfigurable reflectarray to achieve beam scanning by adjusting the phase states of individual elements. One key advantage of this system is its fast imaging speed. However, cost control remains a significant challenge due to the large number of reflectarray elements. Fortunately, the proposed statistical sparse sampling strategy can also be applied to thin the reflectarray and mitigate these cost concerns.
In summary, the statistical sparse sampling strategy proposed in this study can be used to optimize the topologies of electrical scanning imaging and SAR imaging systems for a given target area of interest. In addition to security check, this untrained imaging pipeline can also be applied to other imaging scenarios, such as non-destructive testing.

The paramount objective of the aforementioned MMW imaging techniques for security check is to aid human operators in the detection of concealed hazardous targets. Therefore, we can omit the imaging process and end-to-end extract high-dimensional semantic information of the scene. End-to-end sensing systems\cite{wetzstein2020optically,liu2022programmable} can alleviate storage and bandwidth pressures associated with imaging results. Besides, the end-to-end sensing system remains impervious to signal distortion and loss resultant from imaging generation, thereby facilitating the enhancement of recognition rates. The technique of joint training of the sampling matrix and sensing network, as proposed in the preceding section, finds equal applicability. The aim of this approach is to optimize the sampling matrix for achieving optimal sensing performance, further enhancing the precision and efficiency of concealed target recognition.


\section{Method}\label{sec:method}
\subsection{System implementation}

\begin{figure}[!t]
\centering
\includegraphics[width=1\linewidth]{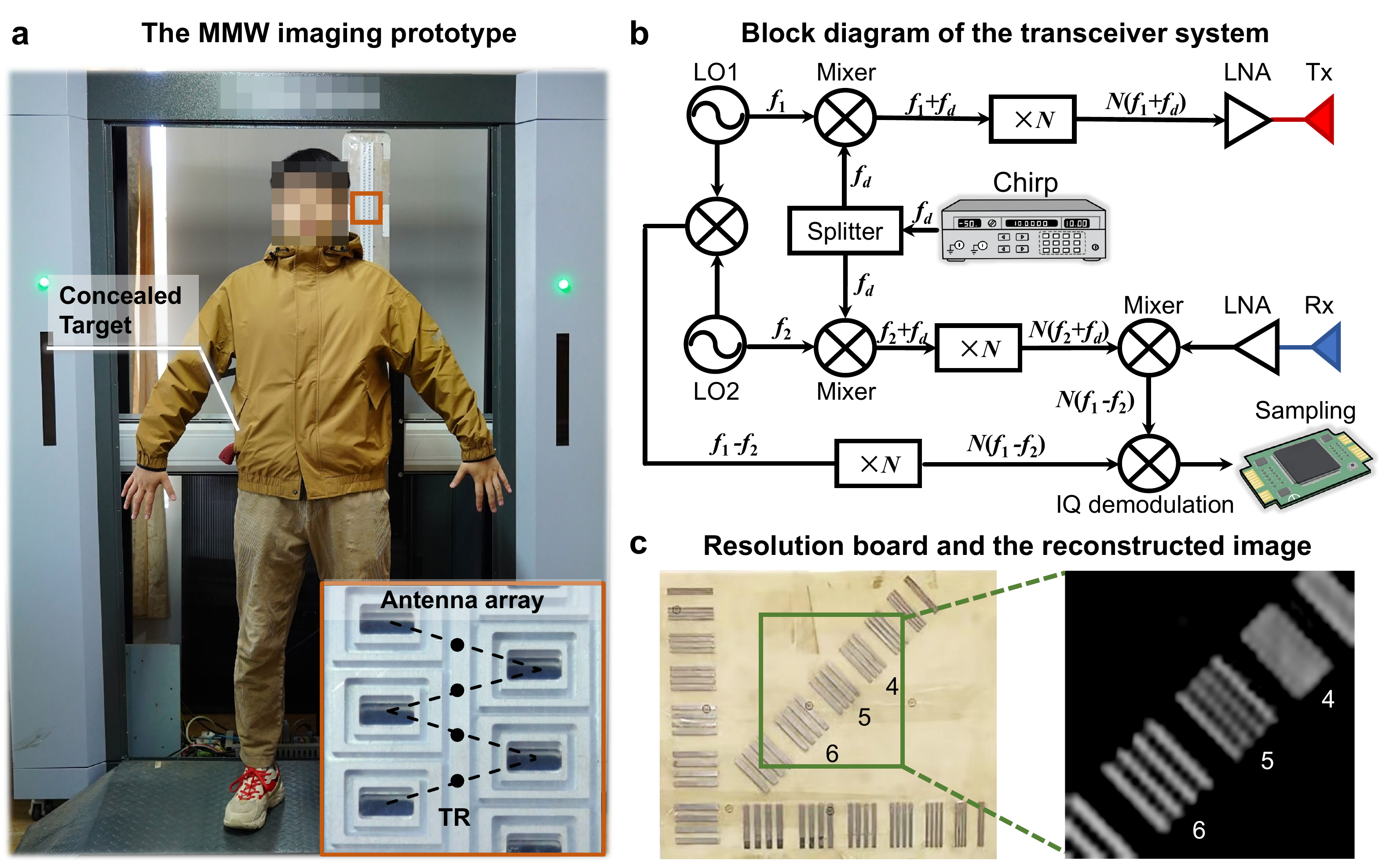}
\caption{The MMW imaging prototype. \textbf{a}, The working system for human security inspection. \textbf{b}, Block diagram of the transceiver system. \textbf{c}, The full-sampled imaging result of a resolution target. The metal bars of Group 4, 5, and 6 on the resolution board correspond to the width and intervals of 4, 5, and 6 mm, respectively.}
\label{fig:system_framework}
\end{figure}

We constructed an MMW imaging prototype for the following experiments. The prototype emits broadband linear frequency modulation (LFM) signal and receives echo with the dechirp technique. The principle diagram of the system is shown in Figure \ref{fig:system_framework} \textbf{b}. The first and second local oscillators (LO1 and LO2)  output carrier signals $f_1$ and $f_2$, respectively, which are mixed with the LFM signal $f_d$ generated by the Direct Digital Frequency Synthesis (DDS), to obtain $f_1 + f_d$ and $f_2 + f_d$, respectively. After $N$-fold frequency multiplication, the transmission signal $N (f_1 + f_d)$ and the reference signal $N (f_2 + f_d)$ are obtained, respectively. The echo $N (f_1 + f_d)$ is de-modulated with the reference signal to obtain the intermediate frequency signal $N (f_1 - f_2)$. The reference intermediate frequency signal $N (f_1 - f_2)$ is obtained by mixing and N-fold frequency multiplication of the LO signals. Finally, the analog IQ demodulation is used to obtain the baseband signal and then digital sampling is performed.

The working frequency of the prototype is from 32 to 37 GHz. The prototype consists of a linear monostatic array with a vertical mechanical scanning structure. The antenna is in the waveguide slot form with vertical linear polarization. The transmit and the receive arrays are arranged in a staggered manner, thus an equivalent sampling point is formed at the center between two neighboring transmit and the receive elements. The sampling interval of the equivalent sampling point meets the system requirements\cite{sheen2010near}.  The transmit antennas work in sequence, with the two neighboring receive antennas collecting the EM wave at the same time. This working pattern leads to 383 virtual monostatic sampling points with an interval of 5 mm. The resolution of the full sampling array is 5mm as illustrated in Fig. \ref{fig:system_framework} \textbf{c}. The linear array scans horizontally to form a 2D array aperture. The scanning spacing is set as 4 mm with 251 moving steps. In this way, an equivalent monostatic full array with  381 $\times$ 251 elements can be achieved. 

\subsection{Analysis on the statistical maps of MMW echoes} \label{sec:method_gradient}

We conducted an analysis on the phase gradient map of a collection of real-captured echoes from a fully-sampled antenna array. This part aims to elucidate the underlying reasons why this map is capable of effectively reflecting the statistical importance ranking of phase in millimeter-wave (MMW) echoes.

We consider a scenario where the target is distributed along the $x$ and $y$ axes. In this context, $\sigma \left(x,y \right)$ represents the scattering coefficient, $k_0$ corresponds to the wavenumber, and $x'$ and $y'$ denote the respective antenna locations along these axes. Additionally, $R_0$ denotes the distance between the target and the antenna array. The gradient along the $x$-axis can be mathematically expressed as follows:
\begin{equation}\label{linear_arr_echo5}
	\begin{aligned}
		\frac{\partial s\left(x',y' \right)} {\partial x'} \!\!= \!\!
		-2C \!\!\int_{y'-y_{ \max}}^{y'+y_{ \max}}  \!\!\!\!\int_{x'-x_{ \max}}^{x'+x_{ \max}} \!\! \sigma \left(x,y \right) e^{-{\rm j}k_0\frac{\left(x'- x\right)^2+\left(y'- y\right)^2}{2R_0}} 
		\!\!\left(x-x'\right) {\rm d}x {\rm d}y,
	\end{aligned}
\end{equation}
where $C$ is a constant number, and 
\begin{equation}
\begin{aligned}
	x_{ \max} =	R_0\tan \frac{\Theta_x}{2},\\
        y_{ \max} =	R_0\tan \frac{\Theta_y}{2}.\\
        \end{aligned}
\end{equation}

Herein, $\Theta_x$ and $\Theta_y$ represent the antenna beamwidth along the azimuth and height directions, respectively. Let us consider a simplified scenario where the scattering coefficient $\sigma (x,y)$ remains constant. Under this assumption, it follows that $\frac{\partial s\left(x',y'\right)} {\partial x'}=0$. Similarly, we can deduce that $\frac{\partial s\left(x',y'\right)} {\partial y'}=0$. Consequently, we can infer that targets with fluctuating scattering coefficients will exhibit a large gradient. In other words, a sparse array with small element gradients indicates better illumination of the target of interest. A more comprehensive analysis is provided in Supplementary Note 1.

Moreover, the averaged amplitude map reflects the significance of the elements within the constraints of the antenna beamwidth and the scope of the subjects. By multiplying the averaged amplitude map with the inverse phase gradient, we obtain a statistical ranking of element importance. This ranking enables us to employ an optimal polynomial sampling strategy for selectively choosing antenna elements at various sampling ratios.

\subsection{Statistically sparse sampling}\label{sec:method_sampling}

We develop a quantitative statistically sparse design to obtain the sparse pattern $M$. Different from handcrafted-designed or other sparse sampling strategies\cite{Efficient2022Li}, we sample the elements in order of statistical importance in $\bar{M}$ with a fixed probability. Given the statistical prior $\bar{M}$ and a uniform random function $r\left(n\right)$ ranging from 0 to 1, we sample the element $n$ of $r\left(n\right) > S$, where $S$ is a hyperparameter to control the sparsity of the sampling pattern. The sampling pattern $M$ can be formulated as
\begin{equation}
    M\left( n \right) =
    \begin{cases}
        1 & r\left( n \right) \geq S, \\
        0 & r\left( n \right) < S, \\
    \end{cases}
\end{equation}
where `1' means the element to be sampled. If the total number of sampled elements is greater than the preset number determined by the sampling ratio, the last out-of-range elements will be discarded.
\subsection{Untrained reconstruction based on CCN}\label{sec:untrained} 

The objective of the reported untrained reconstruction is formulated as
\begin{equation}
\arg \min_{\theta} \| \mathcal{H}\left(f_{\theta} \left( z \right)\right) -  E_{s} \|^2_2 + {\rm TV}\left(f_\theta \left( z \right)\right),
\label{eq:obj}
\end{equation}
where $\theta$ is the parameters of the CCN $f_{\theta}$, $E_s$ is the sparse MMW measurement, $\mathcal{H}$ is the physical model of MMW scattering, and $z$ is the input of the network.
The scattering process $\mathcal{H}$ can be denoted as 
\begin{equation}\label{scattering_process}
\mathcal{H}\left[\cdot \right]=\mathcal{F}_{2D}^{-1}\bigg\{\text{IN}_{k}\Big\{\mathcal{F}_{3D}\left[\cdot \right]\Big\}e^{-\mathrm{j}k_y R_0}\bigg\},
\end{equation}
where $\mathcal{F}_{3D}\left\{\cdot\right\}$ represents a 3-D spatial Fourier transform for all the spatial dimensions of the imaging region. $\mathcal{F}^{-1}_{2D}\left\{\cdot\right\}$ denotes the 2-D spatial inverse Fourier transform over the 2-D array aperture. $\text{IN}_{k}$ indicates the interpolation with respect to the wavenumber $k$. 

As shown in Fig. \ref{fig:network} \textbf{a}, the reported lightweight CCN has 7 blocks, consisting of an input complex Conv-BN-ReLU block, 5 complex Res-blocks, and an output complex Conv layer in sequential. All complex convolutional layers in the network have 256 complex kernels (kernel size=3, step=1, padding=1).
What differentiates CCN from a real-valued convolutional network is the complex convolutional layer. The complex convolution takes the real and imaginary parts of a complex feature as two-channel input and convolutes the input features with complex kernels. We took the reconstructed 3D scene by RMA as the input $z$ and used the Adam\cite{kingma2014adam} solver with a learning rate of 0.001 to update the parameter $\theta$. The network was implemented on the Pytorch 1.13 platform. We take 100 iterations for convergence in most cases.
\bmhead{Complex convolutional layer}
Given the input complex feature map $F = F_R + iF_I$ and the complex convolutional kernel $K = K_R + iK_I$, the complex convolution is denoted as 
\begin{equation}
\begin{split}
F * K &= \left( F_R + iF_I \right) * \left( K_R + iK_I \right) \\
&= \left( F_R*K_R - F_I*K_I \right) + i\left( F_R*K_I + F_I*K_R \right),
\end{split}
\end{equation}
where $*$ denotes the convolution operation. In the implementation, we treat complex values as two-channel real values (real channel $R$ and imaginary channel $I$). 

\bmhead{Complex ReLU}
The complex ReLU ($\mathbb{C}\text{ReLU}$) is denoted as
\begin{equation}
    \mathbb{C}\text{ReLU} = \text{ReLU}(F_R) + i\text{ReLU}(F_I),
\end{equation}
where the ReLU function is
\begin{equation}
    \text{ReLU} \left( F \right) = \left\{
    \begin{aligned}
        &F, \quad \text{if} \  F \geq 0, \\
        &0, \quad \text{otherwise}.
    \end{aligned}
    \right.
\end{equation}

\bmhead{Complex Batch Normalization}
Complex batch normalization ($\mathbb{C}$BN) whitens the complex features by multiplying the 0-centered data ($F-\mathbb{E}(F)$) by the inverse square root of the $2\times2$ covariance matrix $V$
\begin{equation}
    \widetilde{F} = V^{-\frac{1}{2}} \left( F - \mathbb{E}\left(F\right)\right),
\end{equation}
where $\mathbb{E}$ represents the mathematical expectation, and the covariance matrix $V$ is denoted as 
\begin{equation}
    V = \begin{bmatrix} V_{rr} & V_{ri} \\ V_{ir} & V_{ii} \end{bmatrix}
    = \begin{bmatrix}
        \text{Cov} \left( F_R, F_R\right) & \text{Cov} \left( F_R, F_I\right) \\
        \text{Cov} \left( F_I, F_R\right) & \text{Cov} \left( F_I, F_I\right)
    \end{bmatrix}.
\end{equation}
Same as the traditional BN, $\mathbb{C}$BN also has the scaling factors $\gamma$ and $\beta$
\begin{equation}
    \mathbb{C}\text{BN}\left(F\right) = \gamma\widetilde{F}  + \beta.
\end{equation}

\bmhead{Complex Res-block}
The complex res-block contains complex Conv-BN-ReLU-Conv-BN layers arranged sequentially, and the final BN layer’s output feature is directly summed with the input feature. In the experiments, we employ a network architecture comprising five complex Res-blocks.


\backmatter

\bibliography{sn-bibliography}

\bmhead{Acknowledgments}
This work was supported by the National Natural Science Foundation of China under Grants 61971045, 62131003, and 61991451.
\bmhead{Author contributions}
L. B., D. L., S. W., and S. L. conceived the idea. D. L. and S. W. performed the design of the statistical sparse sampling strategy and untrained reconstruction. G. Z and H. S. provided the hardware platform for experiments. D. L., S. W., H. L., C. T., H. X., and R. J. conducted the experiments. L. B., S. L., and J. Z. supervised the project. All the authors participated in the analysis and discussion of the results.
\bmhead{Competing Interests}
L. B. and D. L. hold patents on technologies related to the devices developed in this work (China patent numbers ZL202210778396.7, ZL202010522279.5, and ZL201911081307.8) and submitted related patent applications.

\end{document}